\documentclass[aps,prl,twocolumn,showpacs,amsmath,amssymb,superscriptaddress]{revtex4}

\usepackage{graphicx}
\usepackage{amssymb}
\usepackage{natbib}
\usepackage{calc}
\usepackage{SIunits}
\usepackage{textcomp}
\usepackage{amsmath}
\usepackage{amssymb}
\usepackage{hyperref}
\begin{document}

\title{Neutron Spin Resonance in the Heavily Hole-doped KFe$_{2}$As$_{2}$ Superconductor}

\author{Shoudong Shen}
\thanks{These authors contributed equally to this work.}
\author{Xiaowen Zhang}
\thanks{These authors contributed equally to this work.}
\author{Hongliang Wo}
\author{Yao Shen}
\author{Yu Feng}
\affiliation{State Key Laboratory of Surface Physics and Department of Physics, Fudan University, Shanghai 200433, China}
\author{A.~Schneidewind}
\author{P.~\v{C}erm\'{a}k}
\affiliation{J\"{u}lich Centre for Neutron Science (JCNS) at Heinz Maier-Leibnitz Zentrum (MLZ), Forschungszentrum J\"{u}lich GmbH, Lichtenbergstr. 1, 85748 Garching, Germany}
\author{Wenbin Wang}
\affiliation{Institute of Nanoelectronic Devices and Quantum Computing, Fudan University, Shanghai 200433, China}
\author{Jun Zhao}
\email{zhaoj@fudan.edu.cn}

\affiliation{State Key Laboratory of Surface Physics and Department of Physics, Fudan University, Shanghai 200433, China}
\affiliation{Collaborative Innovation Center of Advanced Microstructures, Nanjing, 210093, China}

\date{\today}

\begin{abstract}
We report high-resolution neutron scattering measurements of the low energy spin fluctuations of KFe$_{2}$As$_{2}$, the end member of the hole-doped Ba$_{1-x}$K$_x$Fe$_2$As$_2$ family with only hole pockets, above and below its superconducting transition temperature $T_c$ ($\sim$ 3.5 K). Our data reveals clear spin fluctuations at the incommensurate wave vector ($0.5\pm\delta$, 0, $L$), ($\delta$ = 0.2)(1-Fe unit cell), which exhibit $L$-modulation peaking at $L=0.5$. Upon cooling to the superconducting state, the incommensurate spin fluctuations gradually open a spin-gap and form a sharp spin resonance mode. The incommensurability ($2\delta$ = 0.4) of the resonance mode ($\sim1.2$ meV) is considerably larger than the previously reported value ($2\delta$ $\approx0.32$) at higher energies ($\ge\sim6$ meV). The determination of the momentum structure of spin fluctuation in the low energy limit allows a direct comparison with the realistic Fermi surface and superconducting gap structure. Our results point to an $s$-wave pairing with a reversed sign between the hole pockets near the zone center in KFe$_{2}$As$_{2}$.
\end{abstract}

\maketitle

In iron pnictide superconductors, it is widely believed that the interband interactions between the hole pockets at the zone center ($\Gamma$) and the electron pockets at the zone edges (M) play an important role in the electron pairing and superconductivity. This is supported by the observation of a resonance mode in the spin fluctuation spectrum at the wave vector ($\pi$, 0) connecting the hole and electron pockets, which is presumably due to the creation of a spin exciton when the superconducting gap function has a sign change between the nested electron and hole Fermi surfaces. This together with the fully gapped superconductivity revealed by various experiments suggests an $s\pm$ wave pairing symmetry in iron pnictides \cite{PCDai}.

With increasing carrier doping from the optimally-doped to the overdoped regime, the shifting of the electronic structure would alter the Fermi surface geometry and superconductivity. In the case of the electron-doped 122 family of iron pnictides, the resonance energy scales with $T_c$ and the resonance spectral weight decreases with growth of the mismatch of the electron (enlarged) and hole (shrunk) Fermi pockets \cite{MiaoyinWang1,KMatan}. The superconductivity and resonance modes disappear in the extremely over electron-doped regime, which is consistent with the $s\pm$ wave pairing scenario associated with the interband interaction.

In hole-doped 122 iron pnictide Ba$_{1-x}$K$_x$Fe$_2$As$_2$, the spin resonance mode also gradually vanishes in the overdoped regime when the mismatch of the hole and electron pockets becomes critical \cite{JPCastellan,CHLee1}. However, surprisingly, accompanied with series of Lifshitz transitions from $x = 0.7$ to $x = 0.9$ \cite{WMalaeb1,NXu}, superconductivity persists up to $x = 1$ where the electron pockets around the zone edges (M) completely disappear. This can not be readily explained by a simple $s\pm$ pairing scenario.

The pairing symmetry in over hole-doped Ba$_{1-x}$K$_x$Fe$_2$As$_2$ has been the subject of intense debate both experimentally and theoretically. In particular, indication of a nodal superconducting gap was revealed by magnetic penetration-depth \cite{KHashimoto,HKim,KCho} and thermal conductivity \cite{JKDong,JPReid,DWatanabe} measurements when the doping level approaches x = 1, which is distinct from the fully gapped superconductivity observed in the optimally-doped and slightly overdoped regime \cite{HDing,LZhao,KNakayama}.
This points to a transition of pairing symmetry near the extremely over hole-doped regime. The thermal conductivity \cite{JPReid} measurements were interpreted as evidence for a $d$-wave pairing symmetry, while angle-resolved photoemission spectroscopy (ARPES) and specific heat experiments argue against this scenario \cite{KOkazaki,Hardy}. In terms of spin fluctuations, although the evolution of the resonance mode over a wide range of doping has been intensively studied in Ba$_{1-x}$K$_x$Fe$_2$As$_2$, the interplay between spin fluctuations and superconductivity in the end member of the hole-doped KFe$_{2}$As$_{2}$ remains unclear\cite{JPCastellan,ADChristianson,ChenglinZhang,CHLee1,CHLee2,MengWang}. There has also been no consensus on the relative sign of the superconducting gap function on different Fermi pockets.

\begin{figure}[t]
\includegraphics[width=0.5\textwidth]{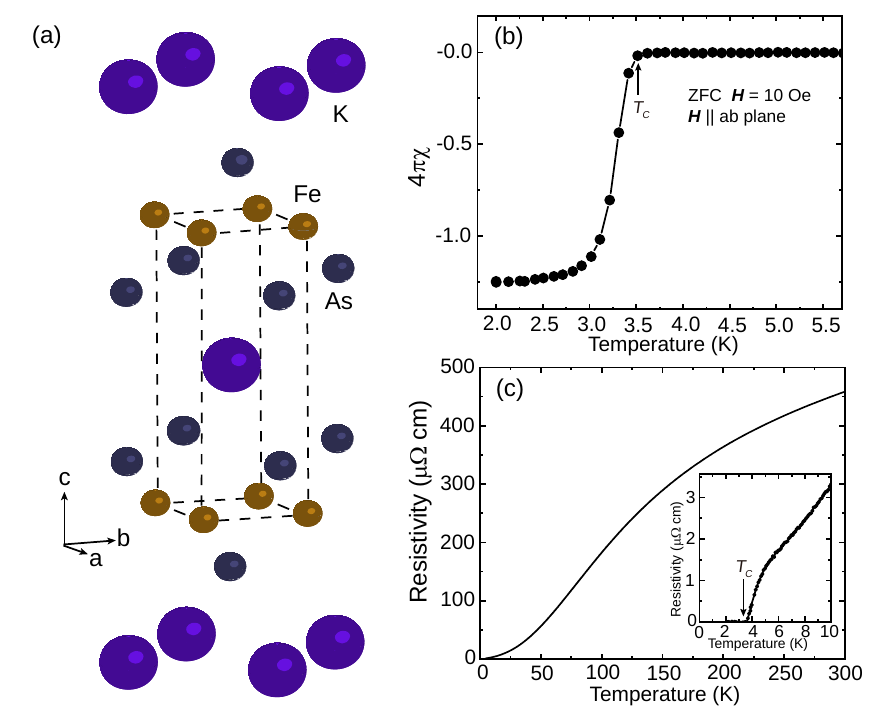}
\caption{
(a) The schematic crystal structure of KFe$_{2}$As$_{2}$. The dashed lines represent the 1-Fe unit cell. (b) DC magnetic susceptibility measurements on the KFe$_{2}$As$_{2}$ single crystal. A sharp superconducting transition is observed at $T_{c}$ $\sim$ 3.5 K in the zero-field-cooled (ZFC) measurement in a magnetic field of $H$ = 10 Oe. The screening is slightly larger than -1 because of the demagnetization effect. (c) Temperature dependence of the in-plane resistivity of KFe$_{2}$As$_{2}$ single crystal. The inset shows data around $T_{c}$ on an enlarged scale.
}
\end{figure}

In this Letter, we report the results of inelastic neutron scattering studies of low energy spin fluctuations in KFe$_{2}$As$_{2}$ single crystals.
We show that the effect of superconductivity is to gradually open a low energy spin gap and also to induce a sharp
resonance mode at energies above the spin gap energy at the incommensurate wave vector ($0.5\pm\delta$, 0, $L$), ($\delta$ = 0.2). The low energy spin fluctuations exhibit $L$-modulation and are peaked at $L=0.5$ in the superconducting state, indicating a non-negligible inter-layer correlation. The resonance mode is also dispersive along the $L$ direction, which is distinct from the nearly two dimensional resonance mode observed in the optimally-doped and slightly overdoped Ba$_{1-x}$K$_x$Fe$_2$As$_2$ \cite{ChenglinZhang,CHLee1}. The incommensurability (2$\delta$ = 0.4) of the low energy spin fluctuation ($\sim1.2$ meV) is considerably larger than that (2$\delta$ $\approx 0.32$) at higher energies($\ge\sim6$ meV) \cite{CHLee2,MengWang,Supplement}. The comparison between the realistic Fermi surface, superconducting gap and low energy spin fluctuation structure implies a sign-reversed pairing gap function between the hole pockets near the zone center in KFe$_{2}$As$_{2}$.

\begin{figure}[t]
\includegraphics[width=0.5\textwidth]{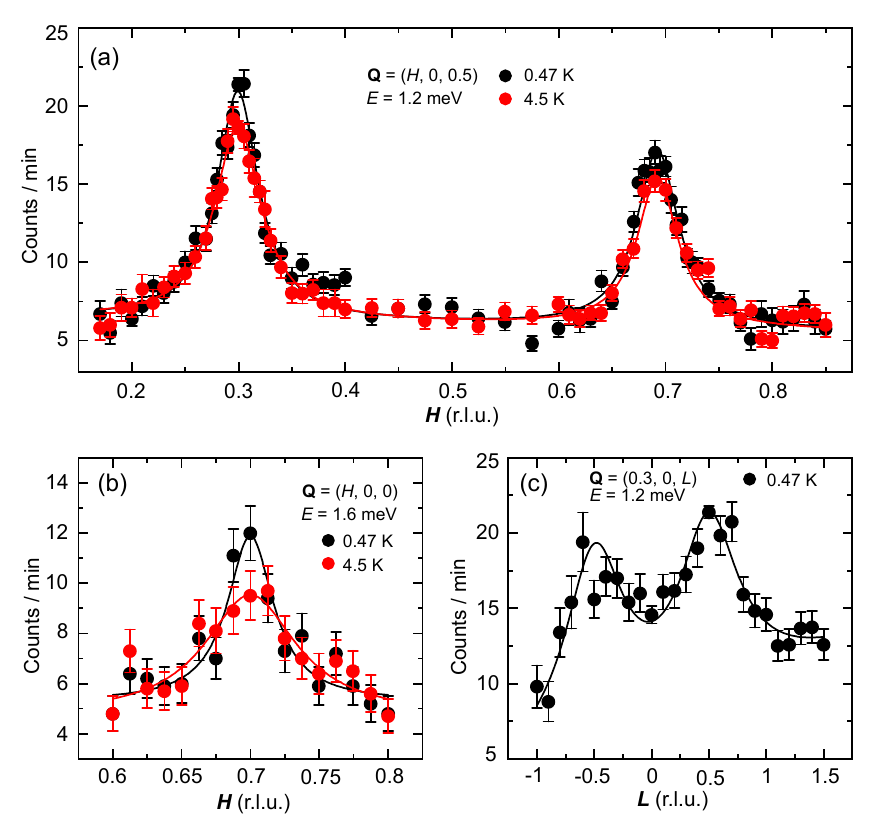}
\caption{
Momentum dependence of the magnetic excitations. (a), (b) Constant energy scans below and above $T_{c}$ at $E$ = 1.2 meV along ($H$, 0, 0.5) direction and at $E$ = 1.6 meV along ($H$, 0, 0) direction, respectively. (c) Constant energy scan along the (0.3, 0, $L$) direction at $E$ = 1.2 meV below $T_{c}$. The spin fluctuations exhibit $L$-modulation and are peaked at the half-integer positions. Solid lines are Lorentz fits to the data.
 }
 \end{figure}

 Single crystals of KFe$_{2}$As$_{2}$ were prepared by the self-flux method similar to that described in Ref.~\onlinecite{AFWang}. K pieces ($\ge$ 99.95\%, Alfa Aesar), Fe ($\ge$ 99.5\%, Sigma-Aldrich) and As ($\ge$ 99.99\%, Alfa Aesar) powders were weighed according to the ratio of K: Fe: As = 4: 2: 5. The mixtures were put into alumina crucibles and then sealed in iron crucibles under an argon atmosphere. The sealed crucibles were heated to 1050 $^{\circ}$C in a tube furnace filled with argon gas, kept at 1050 $^{\circ}$C for 10 hours, and then cooled to 600 $^{\circ}$C in 150 hours. Plate-like single crystals with flat $ab$ surfaces of KFe$_{2}$As$_{2}$ were obtained after the flux were washed out by ethanol.

Magnetic susceptibility and resistivity measurements on pieces from the same batches as the neutron scattering crystals show a sharp superconducting transition at $T_c$ $\sim$ 3.5 K [Fig. 1(b) and (c)], indicating their high quality.
 The neutron-scattering experiments were carried out on the cold triple-axis spectrometer PANDA at the Heinz Maier-Leibnitz Zentrum in M\"{u}nich, Germany. We used double-focusing pyrolytic graphite monochromator and analyzer for all measurements. A cold Be filter was installed before the analyzer and the final neutron energy was fixed at $E_{f}$ = 5.107 meV.
We coaligned 62 pieces of single crystals in the ($H, 0, K$) scattering plane with a total mass of $\sim$ 2.31 g. The sample was loaded in a liquid-He3 orange cryostat to reach the base temperature of 0.47 K. To facilitate the comparison with other iron-based superconductors, we define the wave vector $\textbf{Q}$ in 1-Fe unit cell in reciprocal space as ${\bf Q}=H{\bf a^{*}}+K{\bf b^{*}}+L{\bf c^{*}}$, where $H$, $K$, $L$ are Miller indices and $\textbf{a} ^*=\hat{\textbf{a}}2\pi/a, \textbf{b} ^*=\hat{\textbf{b}}2\pi/b, \textbf{c} ^*=\hat{\textbf{c}}2\pi/c$, with $a = b = 2.709 $ \AA , $c = 6.885 $ \AA.

Fig. 2(a) illustrates constant energy scans at $E$ = 1.2 meV for $L$ = 0.5 at $T$ = 0.47 and 4.5 K.
Well defined incommensurate peaks are observed along the longitudinal direction around ($0.5\pm0.2$, 0, 0.5). We note that the incommensurability $2\delta$ = 0.4 is considerably larger than that ($2\delta$ = 0.32) measured at higher energies $E\ge6$ meV in Ref. \cite{CHLee2}.
The difference could be due to the dispersion of the spin excitation where the incommensurability increases with decreasing energy \cite{MengWang}. Moreover, the scattering intensity at ${\bf Q}$ = (0.7, 0, 0.5) is lower than that at ${\bf Q}$ = (0.3, 0, 0.5), which is roughly in compliance with the Fe$^{2+}$ form factor [Fig. 2(a)]. We also measured the constant energy scans near ${\bf Q}$ = (0.7, 0) at $L$=0 and $\hbar$$\omega$ = 1.6 meV. Clear incommensurate peaks were also revealed with an intensity weaker than that at $L$=0.5, which is inconsistent with an isotropic magnetic form factor. Indeed, the Qscan along the $L$ direction further reveals that the scattering intensity displays a clear $L$-modulation and is peaked at $L=0.5$, which suggests that the inter-layer coupling is antiferromagnetic and non-negligible.
More interestingly, the incommensurate peaks are clearly enhanced below $T_c$ at $0.47$ K, which suggests the presence of superconductivity-induced spectral weight redistribution in KFe$_{2}$As$_{2}$.

\begin{figure}[t]
\includegraphics[width=0.5\textwidth]{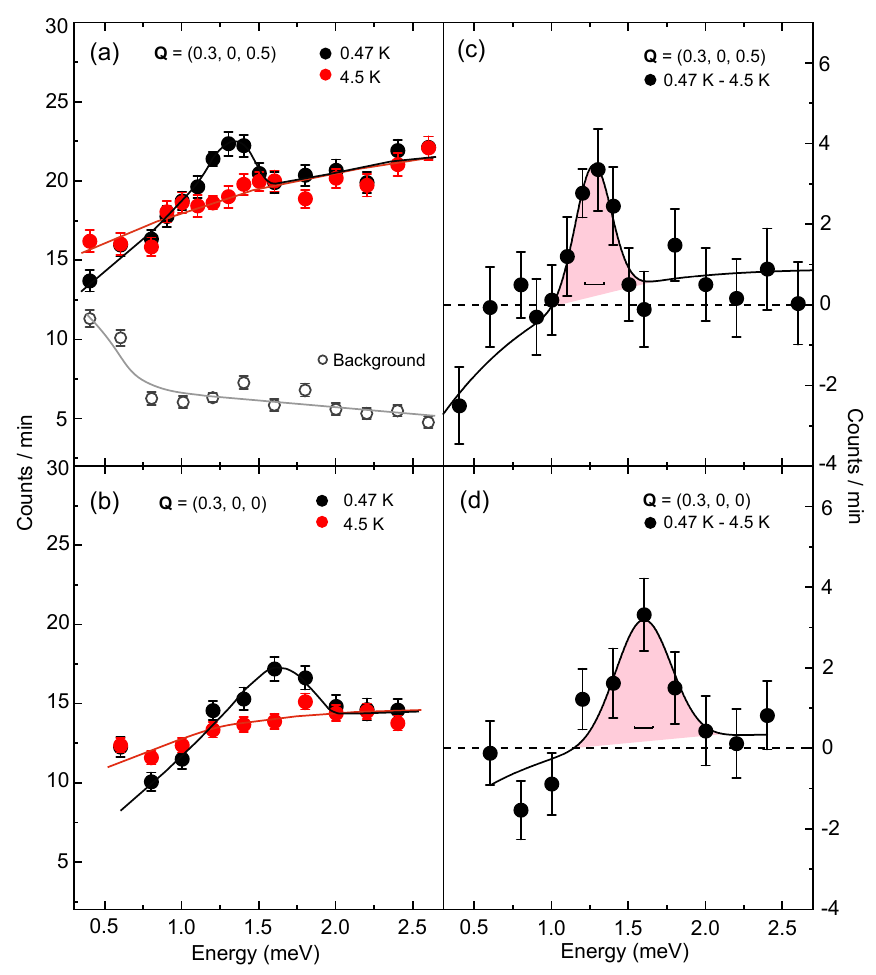}
\caption{
Energy dependence of spin excitations for KFe$_{2}$As$_{2}$ (a),(b) Constant-$\textbf{Q}$ scans in the superconducting state ($\textit{T}$ = 0.47 K) and normal state ($\textit{T}$ =4.5 K) for $\textbf{Q}$ = (0.3, 0, 0.5) and $\textbf{Q}$ = (0.3, 0, 0), respectively. For comparison, the background scattering measured at ${\bf Q}$ = (0.2, 0, 0.5) and (0.5, 0, 0.5) is also shown. (c),(d) Temperature difference plot shows clear spin resonance at $E$ = 1.2 meV for $\textbf{Q}$=(0.3, 0, 0.5) and $E$ = 1.6 meV for $\textbf{Q}$=(0.3, 0, 0), respectively. The horizontal bars indicate the instrument resolution ($\sim$ 0.12 meV). The FWHMs of the resonance peaks at $L$ = 0.5 and $L$ = 0 are $\sim$ 0.27 meV and $\sim$ 0.51 meV, respectively.
}
\end{figure}

To elucidate the origin of the enhanced spin excitations in the superconducting state, we measured the constant-${\bf Q}$ scans at the incommensurate wave vectors ${\bf Q}$ = (0.3, 0, 0.5) and (0.3, 0, 0) above and below $T_{c}$. To eliminate the energy dependent background, the net enhancement is shown by the difference between 0.47 K and 4.5 K data in Fig. 3(c) and Fig. 3(d).
The temperature difference spectrum immediately reveals a sharp resonance mode and a spin gap.
For $L$ = 0.5 [Fig. 3(c)], which is the zone center along the $L$-direction, the resonance is centered at around $E$ =1.2 meV. The full width at half maximum (FWHM) of the resonance peak is $\sim0.27$ meV, which is much sharper than that of other iron pnictide superconductors. This implies that the mode is a spin exciton rather than an enhanced paramagnon due to the loss of particle-hole damping in the superconducting state, which is an important criterion for differentiating sign-reversed and sign-preserved pairing in iron pnictides, though this criterion may not apply in some heavy Fermion superconductors\cite{YSong,TYamashita}.
For $L$ = 0 [Figs. 3(d)], which is the zone boundary along the $L$-direction, the resonance is centered at around $E$ = 1.6 meV.
Both the resonance energy and the spin gap at ${\bf Q}$ = (0.3, 0, 0.5) are lower than that at ${\bf Q}$ = (0.3, 0, 0). This is distinct from the nearly 2-dimensional resonance mode observed in the optimally doped and slightly overdoped Ba$_{1-x}$K$_x$Fe$_2$As$_2$ \cite{ChenglinZhang,CHLee1}. The 3-dimensional nature of the resonance mode may be related to the dispersive behavior of the relevant hole bands along the $L$ direction, as we will discuss subsequently.

To further verify whether the enhanced spin fluctuation in the superconducting state is indeed associated with the superconducting transition, we carefully measured the temperature dependence of the spin fluctuation at ${\bf Q}$ = (0.3, 0, 0.5) and $E=1.2$ meV [Fig. 4(a)].
Indeed, the scattering intensity increases dramatically below the onset of $T_c$, resembling a superconducting order parameter behavior. This is a hallmark of the spin resonance mode, which demonstrates a strong coupling between the appearance of the resonance and superconductivity.

Finally we discuss the origin of the resonance mode and its implication for pairing symmetry in KFe$_2$As$_2$. Since the electron bands do not cross the Fermi level, the hole bands should be active in electron pairing. The Fermi surfaces of KFe$_2$As$_2$ consist of three cylindrical hole pockets [$\alpha$ (inner), $\zeta$ (middle), $\beta$ (outer)] at the zone center $\Gamma$, and four tiny clover-like hole pockets ($\epsilon$) around M [Fig. 4(b)] \cite{KOkazaki,TYoshida,TTerashima}.
It has been shown that the gap is nodeless on the inner pocket, but has eight line nodes on the middle pocket\cite{KOkazaki}. Whether the nodes are symmetry imposed or accidental remains ill-defined.
The outer pocket which mainly consists of $d_{xy}$ orbital shows a negligible gap \cite{OVafek}. Therefore only the inner and middle hole pockets near $\Gamma$ are important for the electron pairing.
Indeed, comparing the low energy spin excitation data and the experimentally determined Fermi surfaces structure reveals that the possible nesting wave vectors between the inner and middle hole pockets near $\Gamma$ are quite close to the incommensurate resonant wave vectors [Fig. 4(b)]. Moreover, in contrast to the rather 2-dimensional outer pocket, the middle pocket shows strong 3-dimensionality \cite{TYoshida}. This is also consistent with the observed 3-dimensional resonance mode.

If the pairing interaction comes from the intraband scattering within the middle pocket, then a $d$-wave pairing seems to be favored. However, a close inspection of the electronic structure reveals that the sign change between the Fermi surface sections nesting along the $H$ direction  within the middle pockets is inconsistent with a conventional $d$-wave symmetry, given that the line nodes are mainly along the diagonal directions on the square-shaped middle pocket \cite{KOkazaki}.
Alternately, if the interband interaction between the inner and middle pockets is dominating, then an $s$-wave pairing with a reversed sign between hole pockets is favored, which is in good agreement with the observed incommensurate resonant peaks at ($0.5\pm0.2$, 0) [Fig. 4(b)].
Several theoretical calculations indeed suggest that the pairing interaction predominantly comes from the inner and middle pockets \cite{OVafek,SMaiti1,SMaiti2,SMaiti3}. We note that the superconductivity induced spectral weight shift of the resonance mode in KFe$_2$As$_2$ is smaller than that of the optimally doped Ba$_{1-x}$K$_x$Fe$_2$As$_2$\cite{ChenglinZhang}. This could be due to the fact that KFe$_2$As$_2$ has a much lower $T_c$ and a much more anisotropic superconducting gap, which may reduce the spectral weight shift in the superconducting state [Fig. 2(a), 2(b)].

\begin{figure}[t]
\includegraphics[width=0.5\textwidth]{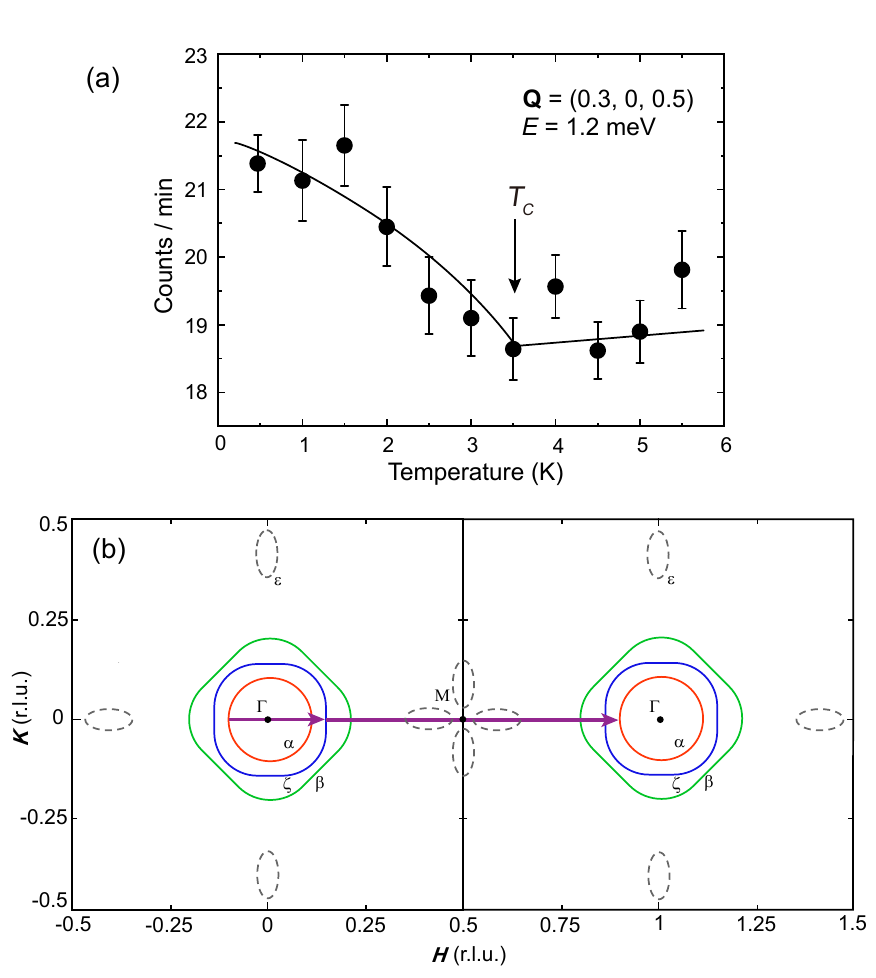}
\caption{
(a) Temperature dependence of the scattering intensity at $\textbf{Q}$ = (0.3, 0, 0.5) with $E$ = 1.2 meV, which clearly shows a kink at $T_{c}$. The data shows clearly order-parameter-like increase below $T_{c}$, a hallmark of the neutron spin resonance. The solid line is a guide to the eyes. (b) Schematic of the hole Fermi pockets in KFe$_2$As$_2$ in the Brillouin zone, inferred from dHvA and ARPES measurements \cite{KOkazaki,TYoshida,TTerashima}.
The purple arrows indicate the interband scattering.
 }
\end{figure}

It has been shown that the resonance energy $E_R$ and $T_c$ follow a universal relation of $E_R$/$k_B$$T_c$ $\approx$ 4.3 in iron pnictide superconductors \cite{JTPark1}. Although KFe$_2$As$_2$ has a much lower $T_c$ and an electronic structure distinct from other iron pnictides, the ratio $E_R$/$k_B$$T_c$ = 4.1 ($L = 0.5$) roughly follows this universal relation. It has also been suggested that $E_R$ and the superconducting gap $2\Delta$ follow an empirical universal ratio of $E_R$/$2\Delta$ $\approx$ 0.64 in unconventional superconductors. However, since the superconducting gap is highly anisotropic with line nodes in KFe$_2$As$_2$, it is difficult to estimate the ratio of the supercondcuting gap and resonance energy. According to ARPES measurements, the maximum of the gap magnitude on the middle and inner pockets is about $\Delta_{max} \approx 0.75$ meV and 1.5 meV at 2 K, respectively \cite{KOkazaki}, which is broadly consistent with scanning tunneling spectroscopy measurements that reveal an average superconducting gap of $\Delta_{avg}$ = 1 meV at a lower temperature of 0.5 K \cite{DFang}. Interestingly, ratio $E_R$/$2\Delta_{avg}$ $\approx$ 0.6 ($L$ = 0.5) is close to the universal ratio of $\sim0.64$ \cite{GYu}. This ratio increases to $E_R$/$2\Delta_{avg}$ $\approx$ 0.8 at $L=0$, which is also similar to the behavior of the 3-dimensional resonance mode observed in the electron-doped and isovalently-doped 122 iron pnictides \cite{SongxueChi,JZhao2013}. The similar characteristics of the resonance mode shared by KFe$_2$As$_2$ and other iron pnictides further imply that they have similar underlying origin.

It is interesting to compare the spin fluctuation spectrum of KFe$_2$As$_2$ with that of the less overdoped Ba$_{1-x}$K$_x$Fe$_2$As$_2$. It has been shown that the spin resonance modes below the superconducting gap in underdoped and optimally doped Ba$_{1-x}$K$_x$Fe$_2$As$_2$ are replaced by a broad hump structure above the superconducting gap at $x\ge0.77$ when the Fermi surface nesting is no longer present due to the vanishing electron pockets. This was interpreted as a reduced impact of magnetism on Cooper pair formation in the overdoped Ba$_{1-x}$K$_x$Fe$_2$As$_2$ \cite{CHLee1}. The reentrance of the incommensurate resonance mode in KFe$_2$As$_2$ is surprising and suggests that a transition of the pairing symmetry may occur between overdoped Ba$_{1-x}$K$_x$Fe$_2$As$_2$ and KFe$_2$As$_2$. This is similar to the behavior of heavily electron doped iron selenide superconductors where the hole pockets disappear and a new resonance mode appears at the incommensurate wave vectors connecting two electron pockets near the zone corner \cite{JTPark2,QWang,BPan}. However, in contrast to the electron doped iron selenide superconductor which shows strong twisted high energy spin fluctuations and has a remarkably high $T_c$ (Ref.~\onlinecite{BPan}), the $T_c$ of KFe$_2$As$_2$ is relatively low and the high energy spin fluctuations ($\ge 20$ meV) are completely suppressed by hole doping \cite{MengWang}. This implies that the pairing symmetry of iron based superconductor is essentially determined by the structures of the low energy spin fluctuations and Fermi surfaces, while the magnitude of $T_c$ is also related to the high energy spin fluctuations.

To conclude, we have performed inelastic neutron scattering experiments to study the low energy spin excitations in KFe$_2$As$_2$ single crystals.
For the first time, we have observed the superconducting spin resonance mode in a heavily hole doped iron based superconductor with no electron pockets.
The resonance mode is dispersive along the $L$ direction, which is consistent with the fact that the hole band shows strong 3-dimensionality. Our results imply an $s$-wave pairing symmetry with a sign-reversed superconducting order parameter between the hole pockets near the zone center.
The persistence of spin resonance mode associated with the hole bands provides a natural explanation for the robustness of superconductivity in KFe$_2$As$_2$ in the extremely overdoped regime.

This work was supported by the Innovation Program of Shanghai Municipal Education Commission (Grant No. 2017-01-07-00-07-E00018), the National Natural Science Foundation of China (Grant No.11874119), the Ministry of Science and Technology of China (Program 973: 2015CB921302) and the National Key R\&D Program of the MOST of China (Grant No. 2016YFA0300203).

\end{document}